\documentclass[twocolumn,prl,floats]{revtex4} 
\usepackage{graphicx,epsfig,amssymb}

\begin{document}
%\draft
%\preprint{}

\title{Signatures of Inelastic Scattering in Coulomb-Blockade Quantum Dots}

\author{T. Rupp, Y. Alhassid, and S. Malhotra} 

\affiliation{Center for Theoretical Physics, Sloane Physics Laboratory, Yale 
University, New Haven, Connecticut 06520}

\begin{abstract}
  We calculate the finite-temperature conductance peak--height
  distributions in Coulomb-blockade quantum dots in the limit where
  the inelastic scattering rate in the dot is large compared with the
  mean elastic tunneling rate. The relative reduction of the standard
  deviation of the peak-height distribution by a time-reversal
  symmetry-breaking magnetic field, which is essentially
  temperature-independent in the elastic limit, is enhanced by the
  inclusion of inelastic scattering at finite temperature. We suggest
  this quantity as an independent experimental probe for inelastic
  scattering in closed dots.
\end{abstract}
%\pacs{ }

\maketitle 

Quantum coherence of the electrons in ballistic quantum dots leads to
distinct signatures in the mesoscopic fluctuations of their
conductance \cite{alhassid00}.  These signatures are affected by
decoherence and inelastic scattering of the electrons at finite
temperature.

Various methods can be used to measure the dephasing rate of the
electrons in open dots \cite{huibers98a,huibers98b}. In particular,
the weak-localization effect (i.e., the reduction of the average
conductance in the absence of a magnetic field) is sensitive to
decoherence, and the suppression of this effect has been used to
extract from experimental data the temperature dependence of the
dephasing time.

In almost-isolated dots, the dephasing and inelastic scattering rates
are expected to vanish at low excitations \cite{sivan94,altshuler97}.
It is, however, more difficult to determine experimentally the
dephasing rate in such dots.  An effect analogous to weak localization
was predicted in the average Coulomb-blockade peak height of a closed
(i.e., almost-isolated) chaotic dot \cite{jalabert92,alhassid98b},
and, in the limit of pure elastic scattering, was found to be
essentially independent of temperature \cite{alhassid98b}. A recent
experiment measured the weak-localization effect in closed dots and
found good agreement with theory for temperatures $kT$ below the mean
level spacing $\Delta$ \cite{folk01}.  However, this effect was found
to be suppressed for $kT \agt \Delta$.  Recently, the opposite limit
where the mean inelastic scattering rate $\Gamma_{\rm in}$ is large
compared with the mean elastic rate $\Gamma_{\rm el}$
\cite{beenakker91} was studied, and the weak-localization effect in
the average peak heights was found to be suppressed with increasing
temperature \cite{beenakker01}.  This suppression was stronger than
what was observed in the experiment, and it was concluded that the
experimental data are consistent with $\Gamma_{\rm in} \alt
\Gamma_{\rm el} \ll \Delta$. Ref.  \cite{eisenberg01} has generalized
the master-equations approach of Ref.~\cite{beenakker91} to study the
effect of inelastic scattering on the average peak height in the
crossover regime $\Gamma_{\rm in} \sim \Gamma_{\rm el}$. This allows
one to determine from the data a mean inelastic scattering rate.
  
Here we study additional signatures of inelastic scattering on the
conductance peak--height statistics in closed Coulomb-blockade quantum
dots.  We calculate the peak-height distributions at finite
temperature in the inelastic limit $\Gamma_{\rm in} \gg \Gamma_{\rm
  el}$, and compare them with the corresponding distributions in the
elastic limit.  In particular, we find that the decrease in the
standard deviation of the conductance peak height upon application of
a magnetic field (measured in units of the peak-height standard
deviation in the presence of a magnetic field) is sensitive to
inelastic scattering.  This reduction in the standard deviation is
temperature-independent in the elastic limit but is enhanced as a
function of temperature in the inelastic limit, and therefore can be
used as an experimental signature of inelastic scattering that is
independent of the weak-localization suppression.

In Ref.~\cite{patel98}, the ratio of the standard deviation of the
peak height to its mean value was measured as a function of
temperature in the presence of a magnetic field and compared with
theory. It was found that the experimental values are suppressed in
comparison with theoretical values calculated in the elastic limit
\cite{alhassid98a}, and the suppression is stronger at higher
temperatures.  To determine whether this suppression can be explained
by inelastic scattering, we calculate the ratio of standard deviation
to mean peak height in the inelastic limit.  While we find some
suppression (compared with the elastic limit), it is relatively small
and insufficient to explain the observed data.

We consider a closed dot that is weakly coupled to the leads, where
the average width $\Gamma_{\rm el}$ for elastic decay of an electron
into the leads is much smaller than the mean spacing $\Delta$ of the
single-particle levels in the dot. We also assume that the thermal
energy $kT$ is much smaller than the charging energy $e^2/C$, where
$C$ is the capacitance of the dot (Coulomb-blockade regime).  To
identify the effect of inelastic scattering of the electrons within
the dot on the linear conductance, we compare two limits. In the
elastic limit, the inelastic decay rate is assumed to be negligible
compared with the elastic rate $\Gamma_{\rm in} \ll \Gamma_{\rm el}$,
and an electron decays into one of the leads prior to any inelastic
scattering event inside the dot. In the opposite limit of strong
inelastic scattering, $\Gamma_{\rm in} \gg \Gamma_{\rm el}$, an
electron tunneling into the dot has sufficient time to thermalize
because of inelastic processes before leaving the dot.

{\it Model.} We further assume the limit $\Gamma_{\rm el}, \Gamma_{\rm
  in} < kT$ of sequential tunneling that can be described by a
master-equations approach \cite{beenakker91}.  In the Coulomb-blockade
regime, the conductance displays sharp peaks as a function of gate
voltage.  The conductance $G$ around a Coulomb-blockade peak can be
written as
\begin{equation}
\label{conductance} 
G(T,\tilde{E}_{\rm F}) = \frac{e^2}{h}\, \frac{\pi \Gamma_{\rm el}}{4kT}\, 
g(T,\tilde{E}_{\rm F}) \;,
\end{equation}
where $g(T,\tilde{E}_{\rm F})$ is a dimensionless conductance at
temperature $kT$ and effective Fermi energy $\tilde E_{\rm F}$ (the
latter can be tuned by varying the gate voltage). For $kT, \Delta \ll
e^2/C$ and in the elastic limit, $g=g_{\rm el}$ can be expressed as a
sum of contributions from several levels $\lambda$ in the dot
\cite{beenakker91} 
\begin{equation} 
g_{\rm el} = \frac{2}{\Gamma_{\rm
      el}} \sum_\lambda w_\lambda \frac{\Gamma^{\rm l}_\lambda
    \Gamma^{\rm r}_\lambda} {\Gamma^{\rm l}_\lambda + \Gamma^{\rm
      r}_\lambda} \;,
\label{elastic_conductance} 
\end{equation}
where $\Gamma_\lambda^{{\rm{l}({\rm r})}}$ is the partial elastic
width of the level $\lambda$ to decay to the left (right) lead, and
$\Gamma_{\rm el} = \overline{\Gamma^{\rm l}_\lambda + \Gamma^{\rm
    r}_\lambda}$.  The contributions of the levels $\lambda$ in Eq.
(\ref{elastic_conductance}) are weighted by factors
\begin{equation}
w_\lambda = 4 f(\Delta F_N - \tilde{E}_{\rm F}) \langle n_\lambda \rangle_N 
[1-f(E_\lambda  - \tilde{E}_{\rm F})]\,, 
\end{equation}
which are a function of both $T$ and ${\tilde E}_{\rm
  F}$~\cite{alhassid98a}.  Here, $f$ is the Fermi-Dirac distribution
function, $\Delta F_N = F_N - F_{N-1}$, where $F_N$ is the canonical
free energy of $N$ noninteracting particles, and $\langle n_\lambda
\rangle_N$ are the canonical occupation numbers of the levels
$\lambda$ in the dot containing $N$ electrons.

On the other hand, in the inelastic limit $\Gamma_{\rm in} \gg
\Gamma_{\rm el}$, there is full thermalization of the electrons among
the levels in the dot and $g=g_{\rm in}$ with \cite{beenakker91}
\begin{equation} 
g_{\rm in} = \frac{2}{\Gamma_{\rm el}}
  \frac{\left(\sum_{\lambda} w_{\lambda} \Gamma^{\rm l}_{\lambda}
    \right) \left(\sum_{\mu} w_{\mu} \Gamma^{\rm r}_{\mu}\right)}
  {\sum_{\nu} w_{\nu} (\Gamma^{\rm l}_{\nu} + \Gamma^{\rm r}_{\nu})}
  \, .
\label{inelastic_conductance}
\end{equation}
Note that the inelastic conductance (\ref{inelastic_conductance}) does
not depend explicitly on $\Gamma_{\rm in}$ (in the inelastic limit).
 
In a chaotic dot (e.g., with irregular shape), the single-particle
levels and wave functions follow random-matrix theory (RMT) statistics
\cite{alhassid00}.  In particular, the partial elastic decay widths
are fluctuating quantities that are independent and distributed
according to a Porter-Thomas law $P_{\rm PT}(\Gamma) \propto
\Gamma^{\beta/2 - 1} \exp(-\beta\Gamma/\Gamma_{\rm el})$, where the
symmetry parameter $\beta$ indicates the presence ($\beta=1$) or
absence ($\beta=2$) of time-reversal symmetry. We assume that all
levels have a common average partial width, i.e.,
$\overline{\Gamma^{{\rm l}({\rm r})}_\lambda} = \overline{\Gamma^{{\rm
      l}({\rm r})}_{\mu}}$ for any pair of levels $\lambda$ and $\mu$,
and that the leads are symmetric: $\overline{\Gamma^{\rm l}_\lambda} =
\overline{\Gamma^{\rm r}_{\lambda}}$.

{\it Method.} We calculate the conductance peak--height distributions
in the inelastic limit and compare them with the corresponding
distributions in the elastic limit.  In an RMT approach, we model the
single-particle spectrum $\{E_\lambda\}$ of the quantum dot by the
spectrum of a matrix that belongs to the corresponding Gaussian
ensemble, i.e., the Gaussian orthogonal ensemble (GOE) for conserved
time-reversal symmetry ($\beta=1)$, and the Gaussian unitary ensemble
(GUE) for broken time-reversal symmetry ($\beta=2)$. The widths
$\{\Gamma^{\rm l}_\lambda; \Gamma^{\rm r}_\lambda\}$ are taken as the
squares of corresponding eigenvector components of the random matrix.
For a given number $N$ of electrons, both the canonical free energy
$F_N$ and the occupation numbers $\langle n_\lambda \rangle_N$ are
calculated exactly using particle-number projection~\cite{alhassid98a}
(projecting from grand canonical partition functions with imaginary
chemical potential). The elastic and inelastic conductance peaks at
temperature $T$ are then calculated as a function of the effective
Fermi energy $\tilde{E}_{\rm F}$ from Eqs.~(\ref{elastic_conductance})
and (\ref{inelastic_conductance}), respectively.  Finally, the height
$g_{\rm max}$ of the corresponding conductance peak at temperature $T$
is calculated by maximizing the function $g(T,\tilde{E}_{\rm F})$ with
respect to $\tilde{E}_{\rm F}$. The (numerical) distribution $P(g_{\rm
  max})$ of the conductance peak heights is obtained by calculating
the peak heights for an ensemble of RMT spectra and corresponding
widths.

\begin{figure}[t!!]
\vspace{3mm}
\centerline{\epsfxsize=0.95\columnwidth \epsffile{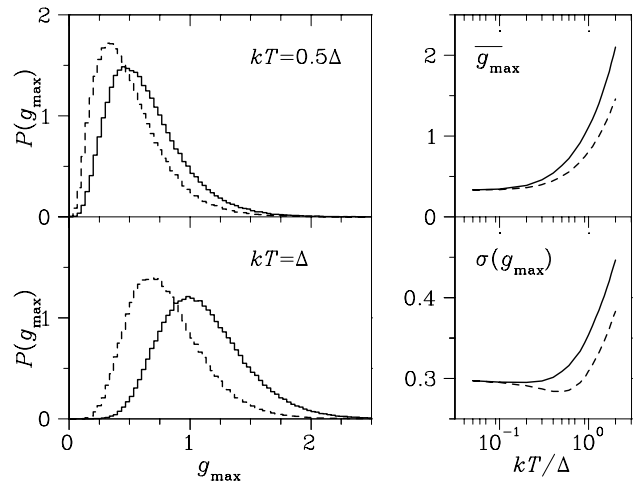}} 
\caption{Left panels: distribution $P(g_{\rm max})$ of the dimensionless peak 
  height $g_{\rm max}$ (see Eq. (\protect\ref{conductance})) in the
  elastic (dashed lines) and inelastic (solid lines) limits at
  temperatures $kT=0.5\Delta$ (top panel) and $kT=\Delta$ (bottom
  panel). Right panels: mean $\overline{g_{\rm max}}$ (top) and
  standard deviation $\sigma(g_{\rm max})$ (bottom) of the peak height
  $g_{\rm max}$ in the elastic (dashed lines) and inelastic (solid
  lines) limits.} \label{fig1}
\end{figure}

{\it Results.}  We have carried out calculations for both
non-degenerate and spin-degenerate single-particle spectra. In both
cases $\Delta$ is the mean-level spacing per single-electron state
(and is therefore half the orbital spacing in the spin-degenerate
case). In the following, we assume non-degenerate energy levels unless
otherwise stated.

Fig.~\ref{fig1} illustrates the temperature dependence of the
distributions $P(g_{\rm max})$ of conductance peak heights $g_{\rm
  max}$ in both the elastic and inelastic limits. In the left panels,
the distributions $P(g_{\rm max})$ are shown at temperatures of $kT =
0.5 \Delta$ and $\Delta$.  In the right panels, the mean conductance
peak height $\overline{g_{\rm max}}$ and standard deviation
$\sigma(g_{\rm max})$($\sigma^2(g_{\rm max}) = \overline{g_{\rm
    max}^2} - \overline{g_{\rm max}}^2$) are shown versus temperature
$kT$. The elastic quantities are shown by dashed lines and the
inelastic quantities by solid lines. All the results in
Fig.~\ref{fig1} are for the case of broken time-reversal symmetry
($\beta=2$), but similar behavior is found for the case of conserved
time-reversal symmetry ($\beta=1$).

In the limit of $kT \ll \Delta$, only a single level $\lambda$
contributes to the various sums in Eqs.~(\ref{elastic_conductance})
and (\ref{inelastic_conductance}) and $g_{\rm el}=g_{\rm in}$. In this
limit, the distributions of the elastic and inelastic conductance
coincide, and the mean peak height and its standard deviation take the
analytically known GUE values for one-level conduction,
$\overline{g_{\rm max}}= 1/3$ and $\sigma(g_{\rm max}) = 2 \sqrt{5}/15
\approx 0.2981$ (the GOE values are $\overline{g_{\rm max}}= 1/4$ and
$\sigma(g_{\rm max}) = \sqrt{2}/4 \approx 0.3536$).  At finite
temperatures, both the mean $\overline{g_{\rm max}}$ and the standard
deviation $\sigma(g_{\rm max})$ are found to be enhanced by inclusion
of inelastic scattering.

\begin{figure}[t!!]
\vspace{3mm}
\centerline{\epsfxsize=0.79\columnwidth \epsffile{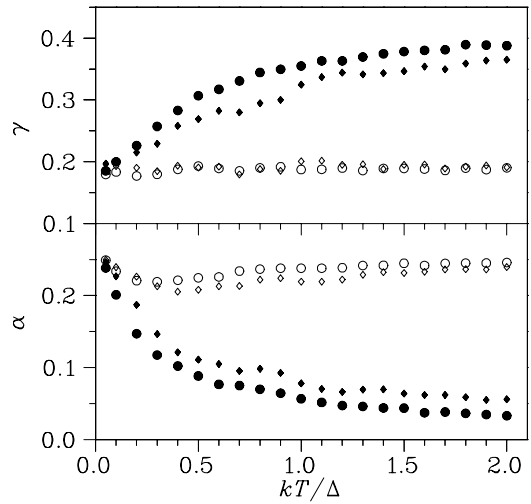}} 
\caption{The  quantities $\gamma$ (upper panel, Eq. (\protect\ref{gamma})) and 
  $\alpha$ (lower panel, Eq. (\protect\ref{alpha})) versus temperature
  $kT$ in the elastic (open circles) and inelastic (solid circles)
  limits. The diamonds are the corresponding results for a
  spin-degenerate spectrum.  Notice the sensitivity of both $\gamma$
  and $\alpha$ to inelastic scattering.}  \label{fig2}
\end{figure}

In recent work~\cite{folk01,beenakker01}, the finite-temperature
suppression of the weak-localization effect in the average conductance
peak height was suggested as a signature of inelastic scattering in
the dot. Quantitatively, the weak-localization effect can be described
by the increase of the average peak height upon breaking of
time-reversal symmetry (relative to the GUE average value)
\begin{equation}
\label{alpha} 
\alpha = 1 - {\overline{g_{\rm
        max}}^{\rm GOE} \over \overline{g_{\rm max}}^{\rm GUE}} \;.
\end{equation}
At low temperatures $kT \ll \Delta$, $\alpha=0.25$
\cite{jalabert92,alhassid98b}.  In the elastic limit, $\alpha$ is
fairly independent of temperature \cite{alhassid98b}, as is shown by
the open symbols in the lower panel of Fig.~\ref{fig2}.  The circles
correspond to a non-degenerate single-particle spectrum and the
diamonds to a spin-degenerate spectrum. As the temperature increases,
more levels contribute to the elastic conductance but the ratio
between the GOE and GUE averages of the peak height remains
approximately constant, since the average over several contributing
levels commutes with the ensemble average.  A small suppression in the
temperature range $ kT/\Delta \alt 0.8$ is due to the suppressed
occurrence of close lying levels in the GUE case, as pointed out
already in Ref.~\cite{eisenberg01}. Because level repulsion is
stronger in the GUE than the GOE, the contribution of an excited level
in the dot to the elastic conductance is smaller on average for the
GUE, and $\alpha$ decreases somewhat.

On the other hand, inelastic scattering is found to reduce $\alpha$ at
finite temperature and $\alpha \to 0$ for $kT\gg \Delta$.  As a
reminder, we show in the lower panel of Fig~\ref{fig2} the temperature
dependence of $\alpha$ in the inelastic limit (solid symbols).  While
the data shown in Ref.~\cite{beenakker01} was obtained assuming
equidistant levels, our results are calculated from RMT spectra
yielding a stronger suppression of the inelastic $\alpha$ at low
temperatures.

Here we suggest an independent signature of inelastic scattering that
is experimentally accessible. We define $\gamma$ as the reduction of
the standard deviation of the conductance peak height (relative to the
GUE standard deviation) when a time-reversal symmetry-breaking
magnetic field is applied, 
\begin{equation}
\label{gamma}
  \gamma={\sigma^{\rm GOE}(g_{\rm max}) \over \sigma^{\rm GUE}(g_{\rm
      max})} - 1 \;.
\end{equation}
In the upper panel of Fig.~\ref{fig2}, we show $\gamma$ versus
temperature for both the elastic (open symbols) and inelastic (solid
symbols) limits.  For $kT \ll \Delta$, $\gamma = 3 \sqrt{5}/(4
\sqrt{2}) - 1 \approx 0.1859$, characteristic for conduction involving
a single level in the dot. In the elastic limit, $\gamma$ is
essentially independent of temperature.  However, at finite
temperature, $\gamma$ is sensitive to inelastic scattering. In the
inelastic limit, $\gamma$ increases from $0.1859$ at $kT \ll \Delta$
to $\approx 0.42 \pm 0.01$ at $kT \gg \Delta$.  The high temperature
limit is obtained using the approximation $w_\lambda \approx
w_\lambda^{(0)}/2$, where $w_\lambda^{(0)} = 1/\cosh^2[(E_\lambda -
\tilde E_F)/2kT]$ are the thermal weights in the absence of charging
energy \cite{alhassid00}.

The quantity $\gamma$ can be calculated in the crossover regime
$\Gamma_{\rm in} \sim \Gamma_{\rm el}$ using the generalized
master-equations approach of Ref.~\cite{eisenberg01}. It can therefore
be used as an independent experimental probe from which the
finite-temperature inelastic scattering rate can be determined.

\begin{figure}[t!]
\vspace{3mm}
\centerline{\epsfxsize=0.90\columnwidth \epsffile{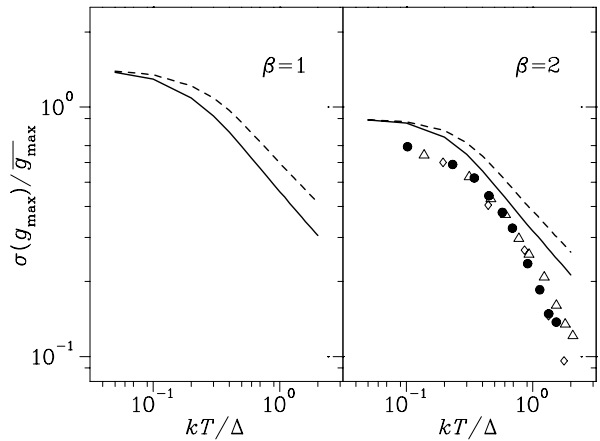}} 
\caption{The  ratio $\sigma(g_{\rm max})/\overline{g_{\rm max}}$ between the 
standard  deviation and average value of the peak height versus  temperature 
$kT$  in the elastic (dashed lines) and inelastic (solid lines) limits for 
both  GOE (left) and GUE (right) symmetries.  The symbols in the right panel 
are  the experimental data of Ref.~\protect\cite{patel98}.} 
\label{fig3}
\end{figure}

Another dimensionless quantity of relevance to the experiments is the
ratio between the standard deviation $\sigma(g_{\rm max})$ and average
$\overline{g_{\rm max}}$ of the peak heights. This ratio was measured
versus temperature in Ref.~\cite{patel98} and found to be
significantly suppressed at higher temperatures in comparison with RMT
calculations \cite{alhassid98a}.  However, these predictions were
based on the assumption of pure elastic scattering, and it was
conjectured that the experimentally observed suppression might be due
to inelastic scattering. Fig.~\ref{fig3} shows the temperature
dependence of $\sigma(g_{\rm max}) / \overline{g_{\rm max}}$ for both
the elastic (dashed lines) and inelastic (solid lines) limits and for
both conserved (left panel) and broken (right panel) time-reversal
symmetry.  The symbols in the right panel are the experimental
results~\cite{patel98}.  While inelastic scattering enhances the
suppression of the ratio $\sigma(g_{\rm max}) / \overline{g_{\rm
    max}}$ at finite temperature, it still cannot explain the
experimental findings at the present stage.  We note, however, that
even at low temperatures there is a discrepancy between the
experimental data and the analytically known value for one-level
conduction ($\sigma(g_{\rm max})/\overline{g_{\rm max}}=
2\sqrt{5}/5\approx 0.8944$ for $\beta=2$). As a probe of inelastic
scattering, the parameters $\alpha$ and $\gamma$ turn out to be far
more sensitive than the ratio between the standard deviation and mean
value. The increases of both the mean and the standard deviation of
the peak height due to inelastic scattering, as seen in the right
panels of Fig.~\ref{fig1}, cancel out to a large extent in the
quantity $\sigma(g_{\rm max}) /\overline{ g_{\rm max}}$, resulting in
a reduced sensitivity of the latter to the inelastic scattering rate.

{\it Conclusion.} We have investigated the effect of inelastic
scattering inside the dot on the distribution of the conductance peak
heights in the Coulomb-blockade regime.  We found that the quantity
$\gamma$ measuring the decrease in the standard deviation of the
conductance peak height upon breaking of time-reversal symmetry
(relative to the value of the standard deviation for broken
time-reversal symmetry) is sensitive to inelastic scattering of
electrons within the dot. This quantity can serve as an independent
experimental probe of inelastic scattering, in addition to the
quantity $\alpha$ that was suggested and measured recently.
 
\begin{acknowledgments}
  We thank C.M. Marcus for useful discussions. This work was supported
  in part by the U.S. DOE grant No.\ DE-FG-0291-ER-40608, and in part
  by the National Science Foundation under Grant No. PHY99-07949.
\end{acknowledgments}

\end{document}